\documentclass[%
groupedaddress,
preprint,
 amsmath,amssymb,
 aps,
prb,
]{revtex4-1}
\usepackage[cp1251]{inputenc}
\usepackage{graphicx}% Include figure files
\usepackage{dcolumn}% Align table columns on decimal point
\usepackage{bm}% bold math

\begin{document}

\title{Lattice source for charge and spin inhomogeneity in 2D perovskite cuprates}
\author{Vladimir A. Gavrichkov}
\email[]{gav@iph.krasn.ru}
%\thanks{}
\affiliation{Kirensky Institute of Physics, Akademgorodok 50, bld.38, Krasnoyarsk, 660036 Russia}
%\affiliation{Siberian Federal University, 660041, Krasnoyarsk, Russia}

\author{Semyon I. Polukeev}
%\thanks{}
\affiliation{Kirensky Institute of Physics, Akademgorodok 50, bld.38, Krasnoyarsk, 660036 Russia}
%\affiliation{Siberian Federal University, 660041, Krasnoyarsk, Russia}

%\author{Sergey G. Ovchinnikov}
%\thanks{}
%\affiliation{Kirensky Institute of Physics, Akademgorodok 50, bld.38, Krasnoyarsk, 660036 Russia}
%\affiliation{Siberian Federal University, 660041, Krasnoyarsk, Russia}
\date{\today}

\begin{abstract}
In the work we highlight the structural features of 2D perovskite cuprates (tilted CuO$_6$ octahedra with different orientation with respect to spacer rocksalt layers), where sources of charge and spin inhomogeneity can be hidden. We used the impurity Anderson model with the Jahn-Teller(JT) local cells to show the charge inhomogeneity arises at any low doping concentration $x$, but disappears when the doping level exceeds threshold concentration $x_c$, and the lower the magnitudes $x_c$, the more JT region square.  It is expected that spontaneous chiral symmetry breaking in the dynamic JT state of the stripe CuO$_2$ layer as a whole can lead to the appearance of the goldstone phonon mode. As consequence, the giant thermal Hall effect could be observed in the 2D perovskite cuprates with CuO$_6$ octahedra, rather than with CuO$_4$ squares, e.g. in Tl-based $n$ layer cuprates or cuprates based on the infinite-layer CaCuO$_2$ structure.
\end{abstract}


\maketitle

\section{\label{sec:intr}Introduction\\}
%\subsection{\label{sec:intr}Introduction\\}

\begin{figure*}
\includegraphics{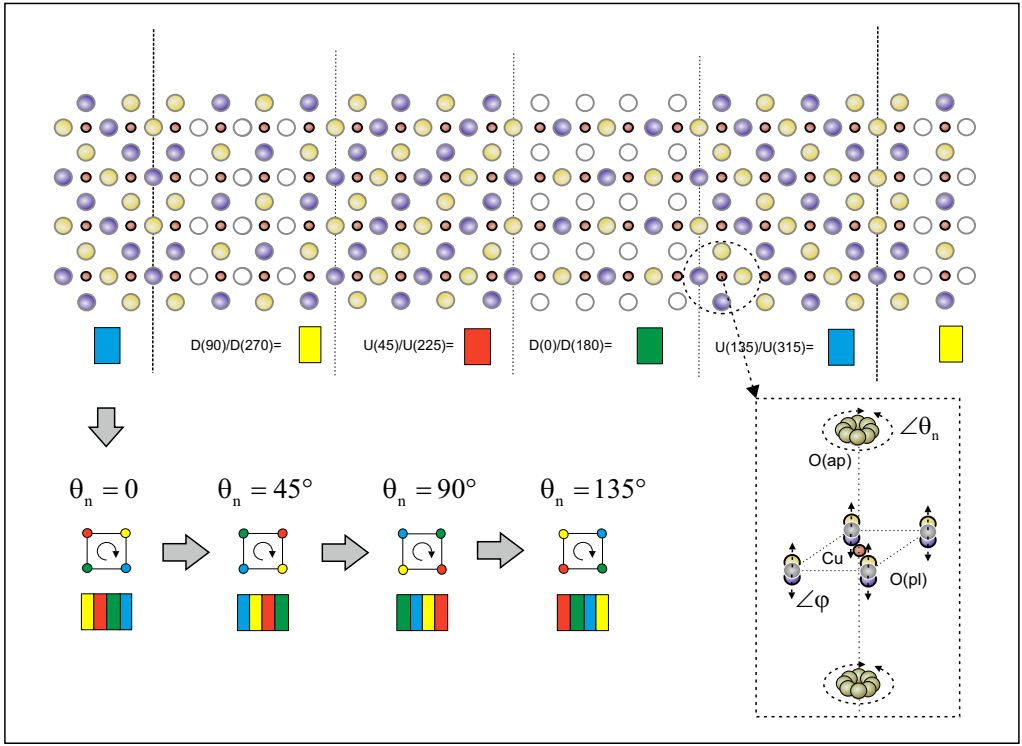}
\caption{The structural motive for the four-fold degenerate $U(\theta)/D(\theta')$ stripe CuO$_2$ layer. The initial phase $\theta_n$ with a color code  and coloring of the planar graph with $\chi=2$ are also shown. The tilted CuO$_6$ octahedron at the tilting angle $\varphi$ and orientation angle $\theta$ in the insert are given for clarification.}
\label{fig:0}
\end{figure*}

In addition to a large isotope effect (see the discussion~\cite{Bussmann2021}), the observed giant thermal Hall effect~\cite{Grissonnanche2019}and its phonon nature \cite{Grissonnanche2020} certainly raises a number of questions about the electronic nature of the charge and spin inhomogeneity of 2D perovskite doped cuprates. In recent studies of the 2D Hubbard model(see, for example, the study~\cite{Robinson2019} and references there), the quasi-degenerate ground states of a Mott-Hubbard material have different physical properties: one of them (A) exhibits long-range order entanglement and translational symmetry breaking, while the other (B) exhibits homogeneous d-wave superconductivity. The difference is confirmed by NMR/NQR experiments.~\cite{Robinson2019} In this electronic scenario a competition between the kinetic energy and Coulomb repulsion can cause holes to segregate into inhomogeneity structures. Studies of the magnetic properties of cuprates  ~\cite{Mahony2022} in contacts with manganites (nicylates) also clearly show magnetic excitations there ~\cite{Deluca2014} (see also the works  ~\cite{Rossat1991, Fong2000, Dai2001} concerning a resonant magnetic mode  within the HTSC phase). There are no phonons in the electronic scenario. On the other hand, using resonant Cu K-edge X-ray diffraction,~\cite{bianconi1996stripe} Cu K-edge X-ray absorption near edge structure (XANES), and Cu K-edge EXAFS,~\cite{bianconi1994instability, missori1994evidence} it was shown that the local structure of the CuO$_2$ layer in the doped perovskite families Bi$_2$Sr$_2$CaCu$_2$O$_{8+y}$(BSCCO)~\cite{bianconi1996stripe, bianconi1994instability, missori1994evidence, saini1998local},  La$_{2-x}$Sr$_x$CuO$_{4+y}$ (LSCO)~\cite{Bianconi1996, lanzara1996temperature, saini2003temperature, saini2003different} show a periodic anharmonic incommensurate lattice modulation. The last is described as a particular nanoscale phase separation due to the formation of alternate stripes of undistorted $U$ stripes and distorted $D$ stripes with different tilts (see Fig.\ref{fig:0}) in the BSCCO and LSCO with the misfit strain between active CuO$_2$ and the spacer Bi2O2 and La2O(2+y) layers.
These results can be analyzed from the symmetry viewpoint, where four subgroups (four colors) are distinguished as the so-called stripe symmetry group $G(\varphi)$ with the tilting angle $\varphi$ of the CuO$_6$ octahedra (Fig.\ref{fig:0}).~\cite{Gavrichkov2019} A spatial structure of the doped CuO$_2$ layer can be constructed in accordance with the well-known four-color theorem~\cite{Biggs_etal1986}, where the chromatic number $\chi  = 2$  corresponds just to the regular stripe structures in Fig.\ref{fig:0})~\cite{Gavrichkov2019, Gavrichkov2022}. The regular structure of the CuO$_2$ layer with the $U(\theta)$ and $D(\theta)$ stripes turns out to be fourfold degenerate with respect to the rotations of all the tilted CuO$_6$ octahedra through the angle $\Delta\theta=45^\circ$ around the $c$ axis in Fig.\ref{fig:0}.
The rotation leads to a uniform shift of linear and chess structures along the transverse and diagonal directions, respectively, but the cell, which has the Jahn-Teller nature,~\cite{Gavrichkov2019, Gavrichkov2022} remains unchanged (see Fig.~\ref{fig:0}). In the dynamic JT effect, the CuO$_2$ layer is in the mixed state, and the orientation angle $\theta$ of the tilted CuO$_6$ octahedra is undefined. This is a chirally symmetric state, and the accompanying anisotropic effects of nonlocality are actually observed in the thermal anisotropic motion as evidenced by neutron experiments with the LCO cuprate.~\cite{Hafliger2014}  Moreover, the spatially homogeneous antiferromagnetic interaction is restored in the hole-doped (rather than electron doped) CuO$_2$ layer in the dynamic stripe state.~\cite{Gavrichkov2022} Coherent tunneling of the CuO$_6$ octahedra is possible in two opposite directions (clockwise and vice versa), and the preferred direction of tunneling of the octahedra corresponds to spontaneous breaking of the chiral symmetry of the CuO$_2$ layer with the tilted octahedra.

\begin{figure*}
\includegraphics{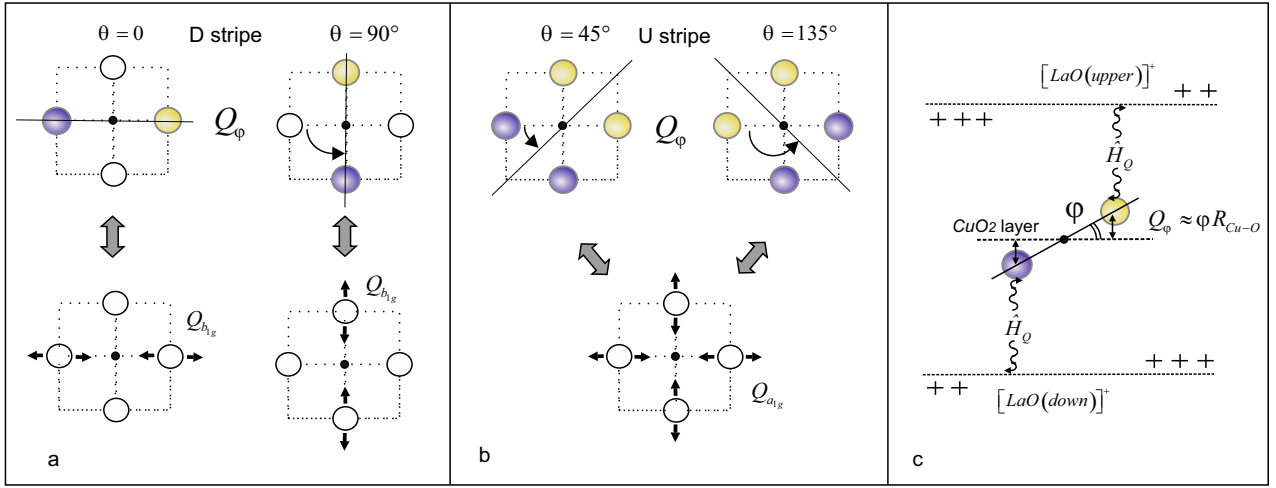}
\caption{The graphic scheme for: (a) b$_{1g}$ tilting modes active in the non-local JT effect of the CuO$_6$ octahedra, and (b) a$_{1g}$ relaxation modes into the D and U stripes, respectively. The arrows in (a,b) show the relationship between the tilting and conventional local modes. (c) Non-local JT effect in the CuO$_2$ layer surrounded by the spacer rocksalt LaO layers.}
\label{fig:1}
\end{figure*}

The key idea here is the JT activity of two out of the four tilts of the CuO$_6$ octahedra in the nonlocal JT pseudo effect under an electrostatic field of the spacer rock salt layers (see Fig.\ref{fig:1}) ~\cite{Gavrichkov2019, Gavrichkov2022}. However, the JT effect itself in 2D Mott-Hubbard materials has specific features in the chemical molecular approach, where the rotational and shift modes are excluded from consideration \cite{Bersuker1989}, and the charge of the JT cell, as a whole, is fixed. In anisotropic 2D perovskite materials, these factors have physical significance. Although the comprehensive study (IR spectroscopy) of the on-molecular JT effect in doped A$_3$C$_{60}$ fullerides reliably detected the metal JT  state,~\cite{Zadik_etal2015} the question on the possibility of the JT (pseudo) effect in doped 2D Mott-Hubbard materials still remains open. Indeed, the JT effect in the  CuO$_2$ layer has the pseudo and  nonlocal nature, where a remarkable feature in the 2D Mott-Hubbard materials consists in its threshold behavior at varying carrier concentrations, for example, under hole doping.

To study the JT effect in a hole doped semiconductor, we use the Anderson impurity model with hybridization and localized magnetic moments, where the undoped semiconductor consists of $N$ cells with pseudo degeneracy in the hole configuration sector (e.g. $\Delta=E_{^1B_{1g}}-E_{^1A_{1g}}\sim J_H$ in LCO, where $J_H$ is a Hund exchange).~\cite{Gavrichkov2022} The JT distortion $Q_\varphi=0$ and there are no signatures of the (pseudo) JT effect, since the hole configuration sector in the undoped semiconductor is empty. With the hole doping $x$ exceeding the threshold values $x_c$, the JT distortion $Q_\varphi$ reaches its maximum magnitudes when the hole is localized in the JT cell. It is a JT  polaron scenario, which can be identified by the quadratic concentration dependence $Q_\varphi(x)\sim x^2$, and instead of localized magnetic impurities, there are JT cells which are active in the JT effect with the localized magnetic moments and distortions $Q_\varphi\neq 0$. However, in contrast to the Anderson model, the number of the impurities $N_{JT}$ is arbitrary and limited only by the total cell number $N$.

\section{\label{sec:II} Non-zero JT distortion in doped semiconductor}
\subsection{\label{sec:JT} JT effect\\}
Let us consider the immersion of the localized (pseudo-)degenerate cell states $|a_\sigma\rangle$ and $|b_\sigma\rangle$ with the energies $\omega_a$ and $\omega_b$ into the valence  band $\varepsilon_{k\sigma}$ of the doped semiconductor material, where the atoms of the immersed JT cell can lose the non-zero JT distortion $Q_\varphi$. How does the value $Q_\varphi$ depend on the hole concentration and  properties of the JT cell and solvent semiconductor material?

The Anderson Hamiltonian describing the JT cell in the semiconductor can be represented as $\hat H = \hat H_{corr}  + \hat H_Q  + \hat H_{mix} $, where:
\begin{widetext}
\begin{eqnarray}
%\begin{array}{l}
&&H_{corr}  = {\textstyle{1 \over 2}}U\left\{ {n_a \sum\limits_\sigma  {\hat n_{a\sigma } }  + n_b \sum\limits_\sigma  {\hat n_{b\sigma } }  + \frac{{2m_a }}{{g\mu _B }}\sum\limits_\sigma  {\eta \left( \sigma  \right)\hat n_{a\sigma } }  + \frac{{2m_c }}{{g\mu _B }}\sum\limits_\sigma  {\eta \left( \sigma  \right)\hat n_{b\sigma } } } \right\} +  \\
\nonumber
&&+ {\textstyle{1 \over 2}}\left( {2U' - J_H } \right)\left\{ {n_b \sum\limits_\sigma  {\hat n_{a\sigma } }  + n_a \sum\limits_\sigma  {\hat n_{b\sigma } } } \right\} + {\textstyle{1 \over 2}}J_H \left\{ {\frac{{2m_b }}{{g\mu _B }}\sum\limits_\sigma  {\eta \left( \sigma  \right)\hat n_{a\sigma } }  + \frac{{2m_a }}{{g\mu _B }}\sum\limits_\sigma  {\eta \left( \sigma  \right)\hat n_{b\sigma } } } \right\}
%\end{array}
\\ \label{eq:1}
&&\hat H_Q  = \sum\limits_\sigma  {\left( {\omega _a a_\sigma ^ +  a_\sigma ^{}  + \omega _b b_\sigma ^ +  b_\sigma ^{} } \right)}  - IQ_\varphi\sum\limits_\sigma  {\left( {b_\sigma ^ +  b_\sigma ^{}  - a_\sigma ^ +  a_\sigma ^{} } \right)}  + \frac{{kQ_\varphi^2 }}{2} \nonumber \\
&&\hat H_{mix}  = \sum\limits_{k\sigma } {\omega _k c_{k\sigma }^ +  c_{k\sigma } }  + \sum\limits_{k\sigma } {\left\{ {V_k c_{k\sigma }^ +  \left( {a_{k\sigma }  + b_{k\sigma } } \right) + V_k^* \left( {a_{k\sigma }^ +   + b_{k\sigma }^ +  } \right)c_{k\sigma } } \right\}}
\nonumber
\end{eqnarray}
\end{widetext}

\begin{figure*}
\includegraphics{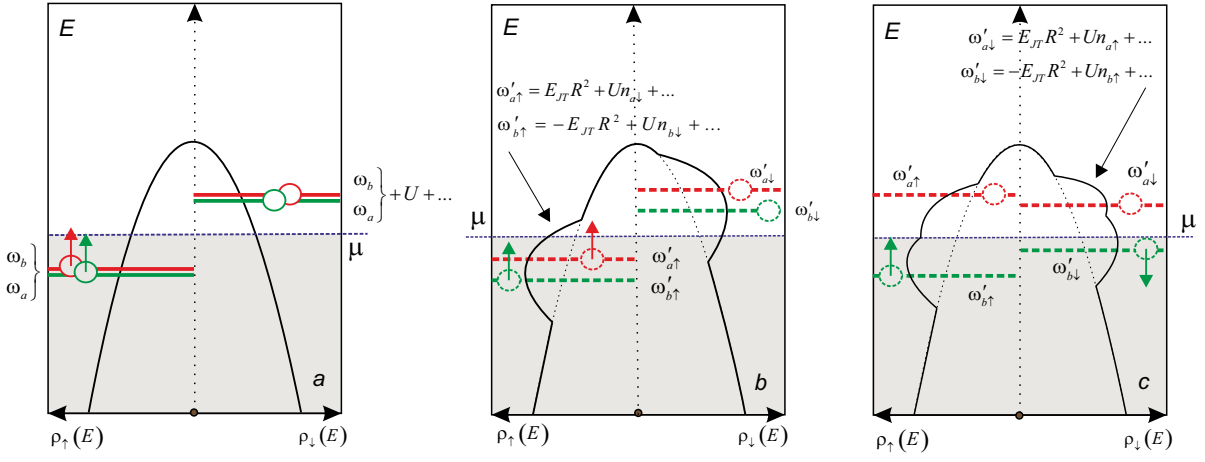}
\caption{(a) The unperturbed energy levels $\omega_a$ and $\omega_b$  at the zero hybridization $V_k=0$ ($n_{\uparrow}=2$, $n_{\downarrow}=0$ and $E_{JT}=0$), the density of the states of the localized levels  is the function $\delta$. The energy levels perturbed by the hybridization $V_k\neq 0$ ($n_{\uparrow}<2$ and $n_{\downarrow}>0$), (b): $U>E_{JT}$ and (c): $U<E_{JT}$}
\label{fig:2}
\end{figure*}
\begin{figure}
\includegraphics{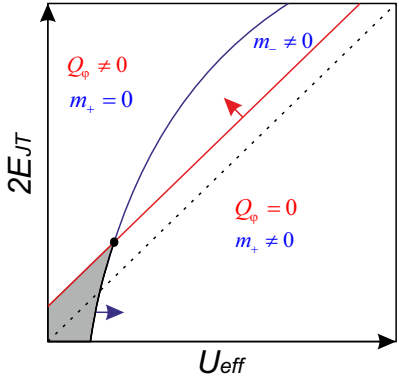}
\caption{The non-zero magnitude of the local magnetic moment and JT distortion regions $Q_ {\varphi}$  upon some mixing $V_k\neq 0$  in the $U$ and $E_{JT}$ coordinates. The arrows show the decreasing region $Q_\varphi\neq 0, m_-\neq 0 $ with the increasing hybridization; the shaded region corresponds to the violation of all the criteria (Eqs.(\ref{eq:6}), (\ref{eq:8}), (\ref{eq:9}), (\ref{eq:13})).}
\label{fig:3}
\end{figure}

We consider the JT (pseudo) effect in the local JT cell with the two-fold  degenerate resonant energy level $\omega _{a,b}$ (see Fig.\ref{fig:2}), where there is non-zero mixing $V_k$ of the band states $c_{k\sigma }$ and localized cell states $|a_\sigma\rangle$ and $|b_\sigma\rangle$.
Here, $H_{corr}$ and $\hat H_{mix}$ are the contributions of the electron correlations in the JT cell in the Hartree-Fock approximation and hybridization of the cell states with the states of the doped parent material. Accordingly,  $n_\lambda   = \sum\limits_\sigma  {\left\langle {n_{\lambda \sigma } } \right\rangle }$ and $m_\lambda   = \frac{{g\mu _{B} }}{2}\sum\limits_\sigma  {\eta \left( \sigma  \right)\left\langle {n_{\lambda \sigma } } \right\rangle }$, $(\lambda  = a,b)$ are  the mean number of electrons and the magnetic moment in the cell states. The parameters  $U$, $U'$, $J_H$, $V_k$ are  the Coulomb interaction in one of the cell states and in different states, the exchange interaction and the hybridization parameter, respectively. At the minimum of adiabatic potential ${{\partial \hat H_Q } \mathord{\left/
 {\vphantom {{\partial \hat H_Q } {\partial Q_\varphi}}} \right.
 \kern-\nulldelimiterspace} {\partial Q_\varphi}} = 0$, $\hat H_Q  =  - IQ_\varphi\sum\limits_\sigma  {\left( {b_\sigma ^ +  b_\sigma ^{}  - a_\sigma ^ +  a_\sigma  } \right)}  =  \pm E_{JT} R\sum\limits_\sigma  {\left( {b_\sigma ^ +  b_\sigma ^{}  - a_\sigma ^ +  a_\sigma  } \right)}$, where $E_{JT}  = {{I^2 } \mathord{\left/
 {\vphantom {{I^2 } k}} \right.
 \kern-\nulldelimiterspace} k}$. The JT distortion $Q_ { \varphi}   =  \frac{IR}{k}$, $R = n_b  - n_a$ and the magnetic moment $m_+ = \sum\limits_\lambda  {m_\lambda  }$ is determined by the electron number per JT cell with the spin $\sigma$: $\left\langle {n_{\lambda \sigma } } \right\rangle  = \int\limits_{ - \infty }^\mu  {\left( { - \frac{1}{\pi }} \right){\mathop{\rm Im}\nolimits} \left\langle {\left\langle {{\lambda _\sigma  }}
 \mathrel{\left | {\vphantom {{\lambda _\sigma  } {\lambda _\sigma ^ +  }}}
 \right. \kern-\nulldelimiterspace}
 {{\lambda _\sigma ^ +  }} \right\rangle } \right\rangle } _{E + i0} dE$, where to calculate the values $R$ and $m_\pm$ we introduce into consideration the anticommutator Green's function $\left\langle {\left\langle {{\lambda _\sigma  }}
 \mathrel{\left | {\vphantom {{\lambda _\sigma  } {\lambda _\sigma ^ +  }}}
 \right. \kern-\nulldelimiterspace}
 {{\lambda _\sigma ^ +  }} \right\rangle } \right\rangle$, for which the equation of motion has the form:
\begin{widetext}
\begin{equation}
\left\{ {E - \omega '_a  - \frac{1}{N}\sum\limits_k {\frac{{V_k^2 }}{{\left( {E - \omega _k } \right)}}} } \right\}\left\langle {\left\langle {{a_\sigma  }}
\mathrel{\left | {\vphantom {{a_\sigma  } {a_\sigma ^ +  }}}
\right. \kern-\nulldelimiterspace}
{{a_\sigma ^ +  }} \right\rangle } \right\rangle  = 1 + \frac{1}{N}\sum\limits_k {\frac{{V_k^2 }}{{\left( {E - \omega _k } \right)}}} \left\langle {\left\langle {{b_\sigma  }}
\mathrel{\left | {\vphantom {{b_\sigma  } {a_\sigma ^ +  }}}
\right. \kern-\nulldelimiterspace}
{{a_\sigma ^ +  }} \right\rangle } \right\rangle,
\label{eq:2}
\end{equation}
\end{widetext}
where
\begin{widetext}
\begin{equation}
\left\{ {E - \omega '_b  - \frac{1}{N}\sum\limits_k {\frac{{V_k^2 }}{{\left( {E - \omega _k } \right)}}} } \right\}\left\langle {\left\langle {{b_\sigma  }}
\mathrel{\left | {\vphantom {{b_\sigma  } {a_\sigma ^ +  }}}
\right. \kern-\nulldelimiterspace}
{{a_\sigma ^ +  }} \right\rangle } \right\rangle  = \frac{1}{N}\sum\limits_k {\frac{{V_k^2 }}{{\left( {E - \omega _k } \right)}}} \left\langle {\left\langle {{a_\sigma  }}
\mathrel{\left | {\vphantom {{a_\sigma  } {a_\sigma ^ +  }}}
\right. \kern-\nulldelimiterspace}
{{a_\sigma ^ +  }} \right\rangle } \right\rangle
\label{eq:3}
\end{equation}
%\end{widetext}
%\begin{widetext}
\begin{equation}
\omega '_\lambda   = \omega _\lambda   - E_{JT} R^2  + \frac{1}{2}U\left\{ {n_\lambda   + \eta \left( \sigma  \right)\frac{{2m_\lambda  }}{{g\mu _B }}} \right\} + \frac{1}{2}\left\{ {\left( {2U' - J_H } \right)n_{\bar \lambda }  + J_H \eta \left( \sigma  \right)\frac{{2m_{\bar \lambda } }}{{g\mu _B }}} \right\}, \lambda  \ne \bar \lambda
\nonumber
\end{equation}
\end{widetext}
Taking into account Eq.(\ref{eq:3}) we have a closed system of equations from which we obtain:
\begin{eqnarray}
&&{\mathop{\rm Im}\nolimits} \left\langle {\left\langle {{a_\sigma  }}
\mathrel{\left | {\vphantom {{a_\sigma  } {a_\sigma ^ +  }}}
\right. \kern-\nulldelimiterspace}
{{a_\sigma ^ +  }} \right\rangle } \right\rangle  = \Gamma \left\{ {\frac{{\alpha _\sigma ^2 }}{{\left( {E - \varepsilon _\sigma ^ +  } \right)^2  + \Gamma ^2 }} + \frac{{\beta _\sigma ^2 }}{{\left( {E - \varepsilon _\sigma ^ -  } \right)^2  + \Gamma ^2 }}} \right\} \nonumber \\
&&{\mathop{\rm Im}\nolimits} \left\langle {\left\langle {{b_\sigma  }}
\mathrel{\left | {\vphantom {{b_\sigma  } {b_\sigma ^ +  }}}
\right. \kern-\nulldelimiterspace}
{{b_\sigma ^ +  }} \right\rangle } \right\rangle  = \Gamma \left\{ {\frac{{\beta _\sigma ^2 }}{{\left( {E - \varepsilon _\sigma ^ +  } \right)^2  + \Gamma ^2 }} + \frac{{\alpha _\sigma ^2 }}{{\left( {E - \varepsilon _\sigma ^ -  } \right)^2  + \Gamma ^2 }}} \right\}, \nonumber \\
\label{eq:4}
\end{eqnarray}
where $\alpha _\sigma ^2  = \frac{1}{2}\left\{ {1 + \frac{{\omega '_a  - \omega '_b }}{{\nu _\sigma  }}} \right\}$, $\beta _\sigma ^2  = \frac{1}{2}\left\{ {1 - \frac{{\omega '_a  - \omega '_b }}{{\nu _\sigma  }}} \right\}$ are the probabilities for the $|\lambda\rangle$ states  after mixing with the valence band states. According to Eq.(\ref{eq:4}), we obtain an equation for the non-zero magnetic moment $m_{+}$ in the JT cell:
\begin{widetext}
\begin{equation}
m_ +   =  - \frac{{g\mu _B }}{2}\sum\limits_\sigma  {\frac{{\eta \left( \sigma  \right)}}{\pi }\int\limits_{ - \infty }^\mu  {dE\left\{ {\frac{\Gamma }{{\left( {E - \varepsilon _\sigma ^ +  } \right)^2  + \Gamma ^2 }} + \frac{\Gamma }{{\left( {E - \varepsilon _\sigma ^ -  } \right) + \Gamma ^2 }}} \right\}} },
\label{eq:5}
\end{equation}
\end{widetext}
where $\varepsilon _\sigma ^ +$ and $\varepsilon _\sigma ^ -$ are the energy levels of the electron in one of the final mixed states, $\Gamma  \sim \pi \left\langle {V^2 } \right\rangle \nu _v \left( \mu  \right)$ and $\nu _v \left( \mu  \right)$ is the density of the states in the valence band. Eq.(\ref{eq:5}) has a solution $m_{+} \ne 0$  at
\begin{equation}
\left( {U + J_H } \right)\left. \rho  \right|_{m_+ = 0} \left( {\mu ,R} \right) > 1
\label{eq:6}
\end{equation}
where  $\rho \left( {\mu ,R} \right) = \rho _a \left( {\mu ,R} \right) + \rho _b \left( {\mu ,R} \right)$ is the density of the JT cell states at the Fermi level $\mu$. This result is similar to the well-known condition for the non-zero localized magnetic moment in a metal~\cite{Anderson1961}. However, the requirements on the value $\left( {U + J_H } \right)$ become more stringent due to the non-zero JT distortion $Q_ { \varphi}$ with a decreasing cell density of the states $\left. \rho  \right|_{m = 0} \left( {\mu ,R\neq 0} \right)$ at the Fermi level $\mu$ . To make sure that the JT distortion $Q_ { \varphi}$  has a non-zero magnitude, we can obtain a condition on the magnitude of the interactions. Indeed, the equation:
\begin{widetext}
\begin{equation}
Q_ { \varphi \pm}   =  \pm \frac{I}{k}\sum\limits_\sigma  {\left( {\alpha _a^2  - \beta _b^2 } \right)\int\limits_{ - \infty }^\mu  {dE\left\{ {\frac{\Gamma }{{\left( {E - \varepsilon _\sigma ^ +  } \right)^2  + \Gamma ^2 }} - \frac{\Gamma }{{\left( {E - \varepsilon _\sigma ^ -  } \right) + \Gamma ^2 }}} \right\}} }
\label{eq:7}
\end{equation}
\end{widetext}
has a solution $Q_ { \varphi }   \ne 0$ in the region $m_ +  =m_a+m_b \ne 0$ only under the condition
\begin{equation}
\left( {2E_{JT}  - U_{eff} } \right)\left. {\sum\limits_\sigma  {\left( {\alpha _\sigma ^2  - \beta _\sigma ^2 } \right)^2 \rho _\sigma ^{} \left( {\mu ,m_ +  } \right)} } \right|_{R = 0}  > 1
\label{eq:8}
\end{equation}
where $\frac{{\partial \varepsilon _\sigma ^ \pm  }}{{\partial R}} =  \pm \left( {\alpha _\sigma ^2  - \beta _\sigma ^2 } \right)\left( {2E_{JT}  - U_{eff} } \right)$ and $U_{eff}  = \frac{1}{2}\left( {2U' - U - J_H } \right) \approx \frac{1}{2}U' - J_H$ and $\rho _\sigma  \left( {\mu ,m_ +  } \right) = \rho _a^\sigma   + \rho _b^\sigma$. Inequality (\ref{eq:8}) is not correct without the JT effect  $E_{JT}  = 0$ and $\left. {\left( {\alpha _\sigma ^2  - \beta _\sigma ^2 } \right)} \right|_{R = 0}  = 0$ at $m_ -= m_a  - m_b  = 0$. Indeed, the condition for the non-zero magnitude $m_ -   \ne 0$   has the form:
\begin{equation}
\left( {U + J_H } \right)\left. {\sum\limits_\sigma  {\left( {\alpha _\sigma ^2  - \beta _\sigma ^2 } \right)^2 \rho _\sigma ^{} \left( {\mu ,m_ +  } \right)} } \right|_{m_ -   = 0}  > 1,
\label{eq:9}
\end{equation}
 where $\left. {\left( {\alpha _\sigma ^2  - \beta _\sigma ^2 } \right)} \right|_{m_ -   = 0}  \ne 0$,  at $R \ne 0$. Thus, the finite magnitude  $m_ -   \ne 0$ can be observed along with    $Q_ { \varphi}   \ne 0$, and there is a zero threshold concentration $x_c  = 0$ in the JT effect. To clarify this, we show in Fig.\ref{fig:3} the region of solutions for the non-zero magnetic moment $m_{\pm} \ne 0$, where the overlapping regions $m_- \ne 0$ and $R \ne 0$  are observed.

Note that the set of inequalities (6) and (8)  does not detect  any non-zero   JT distortion in the region   $Q_ {\varphi}   \ne 0$, where the criterion differs from Eq.(\ref{eq:8}) by non-zero contributions at $m_ +   = m_ -   = 0$.
%and can be obtained by complete differentiation of Eq.(\ref{eq:7}).

\subsection{\label{sec:JT} JT pseudo effect\\}

In the JT pseudo effect, the dependence of the JT distortion on the concentration $Q_ {\varphi \pm}  \left( R \right) =  \pm \sqrt {\left( {\frac{I}{k}R} \right)^2  - \left( {\frac{\Delta }{I}} \right)^2 }$, where $Q_ {\varphi \pm}  = 0$ at the doping concentration $x_c \left( {R_c } \right)$, and $R_c  = {{k\Delta } \mathord{\left/
 {\vphantom {{k\Delta } {I^2 }}} \right.
 \kern-\nulldelimiterspace} {I^2 }}$ has the threshold nature ~\cite{Gavrichkov2022}, and the condition for their non-zero magnitudes in  the equation $Q_ { \varphi }   = Q_ {\varphi }  \left( R \right)$  becomes
\begin{equation}
\left. {\frac{{dQ_\varphi\left( R \right)}}{{dQ_\varphi}}} \right|_{Q_\varphi \to 0}  = \left. {\frac{{\partial Q_\varphi\left( R \right)}}{{\partial R}}} \right|_{R \to R_c }  \times \left. {\frac{{\partial R}}{{\partial Q_\varphi}}} \right|_{Q_\varphi \to 0}  > 1
\label{eq:10}
\end{equation}
In the last inequality  $\left. {\frac{{\partial Q_\varphi\left( R \right)}}{{\partial Q_\varphi}}} \right|_{Q_\varphi \to 0}  \approx \left. {\left( {\frac{I}{k}} \right)^2 \frac{{R_c }}{Q_\varphi}} \right|_{Q_\varphi \to 0}$,  and the derivative  $\left. {\frac{{\partial R}}{{\partial Q_\varphi}}} \right|_{Q_\varphi \to 0}$ is calculated similarly to Eq.(\ref{eq:8}), taking into account the fact that:
\begin{widetext}
\begin{eqnarray}
\omega '_\lambda   = \omega _\lambda(Q_\varphi) + \frac{1}{2}U\left\{ {n_\lambda   + \eta \left( \sigma  \right)\frac{{2m_\lambda  }}{{g\mu _B }}} \right\} + \frac{1}{2}\left\{ {\left( {2U' - J_H } \right)n_{\bar \lambda }  + J_H \eta \left( \sigma  \right)\frac{{2m_{\bar \lambda } }}{{g\mu _B }}} \right\}, \lambda  \ne \bar \lambda
%&&\omega '_a  = \omega _a \left( Q_\varphi \right) + \frac{1}{2}U\left\{ {n_a  + \eta \left( \sigma  \right)\frac{{2m_a }}{{g\mu _B }}} \right\} + \frac{1}{2}\left\{ {\left( {2U' - J_H } \right)n_b  + J_H \eta \left( \sigma  \right)\frac{{2m_b }}{{g\mu _B }}} \right\} \\ \nonumber
%&&\omega '_b  = \omega _b \left( Q_\varphi \right) + \frac{1}{2}U\left\{ {n_b  + \eta \left( \sigma  \right)\frac{{2m_b }}{{g\mu _B }}} \right\} + \frac{1}{2}\left\{ {\left( {2U' - J_H } \right)n_a  + J_H \eta \left( \sigma  \right)\frac{{2m_a }}{{g\mu _B }}} \right\},
\label{eq:11}
\end{eqnarray}
\end{widetext}
where $\omega _{a\left( b \right)} \left( Q_\varphi \right) = \frac{1}{2}\left\{ {\omega _a  + \omega _b  \pm \sqrt {\left( {\omega _a  - \omega _b } \right)^2  + 4\left( {IQ_\varphi} \right)^2 } } \right\}$. As a consequence, the corresponding derivative takes the form:
\begin{widetext}
\begin{equation}
\left. {\frac{{\partial R}}{{\partial Q_\varphi}}} \right|_{Q_\varphi \to 0}  \approx \frac{{kQ_\varphi}}{\Delta }\left( {2E_{JT}  - U_{eff} } \right)\left. {\sum\limits_\sigma  {\left( {\alpha _\sigma ^2  - \beta _\sigma ^2 } \right)^2 \rho _\sigma ^{} \left( {\mu ,m_ +  } \right)} } \right|_{Q_\varphi \to 0}^{\Delta  \ne 0}
\label{eq:12}
\end{equation}
\end{widetext}
and, therefore, in the same way as in the usual JT effect, but at $R \to R_c$, the criterion for the non-zero distortion $Q_ {\varphi} \neq 0$ has the form:
\begin{equation}
\left( {2E_{JT}  - U_{eff} } \right)\left. {\sum\limits_\sigma  {\left( {\alpha _\sigma ^2  - \beta _\sigma ^2 } \right)^2 \rho _\sigma ^{} \left( {\mu ,m_ +  } \right)} } \right|_{R \to R_c }^{\Delta  \ne 0}  > 1
\label{eq:13}
\end{equation}
However, the result, in contrast to Eq.(\ref{eq:8}), has a threshold character at the doping concentration $x > x\left( {R_c } \right)$, where  $R_c  = {{k\Delta } \mathord{\left/
 {\vphantom {{k\Delta } {I^2 }}} \right.
 \kern-\nulldelimiterspace} {I^2 }} \to 0$ at $\Delta  \to 0$. Thus, the diagram in Fig.\ref{fig:2} for the JT pseudo effect does not change. The result is not obvious, since, unlike the previous consideration in terms of the JT effect, the derivative $\left. {\frac{{\partial Q_\varphi\left( R \right)}}{{\partial Q_\varphi}}} \right|_{Q_\varphi \to 0}\rightarrow \infty$.

Here, both in the case of the JT effect and pseudo JT effect, on the $(U,E_{JT})$ phase diagram it is possible to detect a region with the non-zero values $m_-, Q_\varphi$ (see Eqs.(\ref{eq:8}), (\ref{eq:13})), which decreases with an increase in the hybridization $V_k$ of the band and JT cell states. However, the charge inhomogeneity occurs only in the case of the pseudo JT effect.

\section{\label{sec:IV} Discussion and conclusions  \\}

As a result of the JT pseudo effect, JT regions of the lattice are formed with the carrier concentration $x_c \sim x\left({R_c} \right)$ at any low initial concentration $x < x\left({R_c}\right)$, and whose total area grows linearly with the carrier concentration $x$.  The charge inhomogeneity is accompanied by the JT distortions $Q_\varphi(R_c)$, and the lower the threshold magnitudes $x_c$, the larger are the JT regions. In the 2D perovskite cuprates with the $D(\theta)$ stripes, the latter are CuO$_6$  octahedra with the tilting angles $\sim 14^\circ  \div 18^\circ$.~\cite{Bianconi1996} The calculation agrees with the exchange interaction in CuO$_2$ under the JT effect, due to the combined region $m_-\neq0$, $Q_\varphi\neq0$ with the non-zero magnetic moment and JT distortion in Fig.\ref{fig:3}. In the JT materials, where  $x_c=x\left({R_c} \right)=0$, such as 3D perovskites AMeO$_3$ with regular MeO$_6$ octahedra (e.g. in the perovskite oxides LaMnO$_3$, where Mn$^{3+}$ is a JT ion~\cite{Dessau_etal1998}), the hole segregation into dynamic stripe structures is impossible and occurs according to other scenarios~\cite{Kagan_etal2001}. There is no reason to discuss the JT nature of the charge inhomogeneity in 2D n-type LNCO cuprates, since the doped electrons are in the completely occupied 3d shell state and no (pseudo) JT effect is possible there.

We also believe that spontaneous chiral symmetry breaking in the dynamic JT state of all over the CuO$_2$ layer can lead to the goldstone phonon mode.~\cite{Goldstone_1961} However, only apical oxygen performs real rotational motion (see Fig.\ref{fig:0}). As a consequence, the giant thermal Hall effect~\cite{Grissonnanche2020, Grissonnanche2020}  could be observed in HTSC cuprates with CuO$_6$ octahedra, rather than with CuO$_4$ squares, e.g. in Tl-based $n$  layer cuprates Tl$_2$Ba$_2$Ca$_{n-1}$Cu$_n$O$_{2n+4}$,  TlBa$_2$Ca$_{n-1}$Cu$_n$O$_{2n+3}$ or cuprates based on the infinite-layer CaCuO$_2$ structure. Note, the phonon chirality is currently derived from  weak ferromagnetism induced by a strong coupling of the Dzyaloshinsky-Moriya interactions with a specific soft phonon mode in the CuO$_2$ layers .~\cite{Hu_etal2023, Sapkota_etal2021}

The results do not contradict the general structural features observed in perovskite materials,~\cite{Glazer2011, Bersuker_etal2020} and last conclusion  could be verified.

\begin{acknowledgments}
We acknowledge the support of the Russian Science Foundation through grant RSF No.22-22-00298.
\end{acknowledgments}

%\bibliography{PRB_2019}
\bibliography{paper_gavrichkov}
\end{document}